\RequirePackage{lineno}
\documentclass[aps, twocolumn, prl, longbibliography]{revtex4-1}
\usepackage{mathptmx} 
\usepackage{amsfonts}
\usepackage{amsmath}
\usepackage{amssymb}
\usepackage{graphicx}
\usepackage{bm}
\usepackage{natbib}
\usepackage{color}
\usepackage[colorlinks=true,linkcolor=blue,anchorcolor=red,citecolor=blue, urlcolor=blue]{hyperref}

\begin{document}

\title{Pseudospin Berry phase as a signature of nontrivial band topology in a two-dimensional system}
\author{F. Cou\"{e}do}
\altaffiliation[Present address: ]{Laboratoire National de Metrologie et d'Essais (LNE), Quantum electrical metrology department, Avenue Roger Hennequin, 78197 Trappes, France}\thanks{These authors contributed equally to this work.}
\author{H. Irie}\thanks{These authors contributed equally to this work.}
\author{T. Akiho}
\author{K. Suzuki}\altaffiliation[Present address: ]{Fukuoka Institute of Technology, Fukuoka 811-0295, Japan}
\author{K. Onomitsu}
\author{K. Muraki}\thanks{Corresponding author (e-mail: koji.muraki.mn@hco.ntt.co.jp)}
\affiliation{NTT Basic Research Laboratories, NTT Corporation, 3-1 Morinosato-Wakamiya,
Atsugi 243-0198, Japan}
\pacs{PACS number}
\date{\today}

\begin{abstract}
Electron motion in crystals is governed by the coupling between crystal momentum and internal degrees of freedom such as spin implicit in the band structure.
The description of this coupling in terms of a momentum-dependent effective field and the resultant Berry phase~\cite{Xiao2010} has greatly advanced our understanding of diverse phenomena including various Hall effects and has lead to the discovery of new states of matter exemplified by topological insulators~\cite{Qi2011a}.
While experimental studies on topological systems have focused on the gapless states that emerge at the surfaces or edges, the underlying nontrivial topology in the bulk has not been manifested.
Here we report the observation of Berry's phase in magneto-oscillations and quantum Hall effects of a coupled electron-hole system hosted in quantum wells with inverted bands.
In contrast to massless Dirac fermions in graphene, for which the Berry phase $\Gamma$ is quantized at $\pi$~\cite{Novoselov2005,Zhang2005}, we observe that $\Gamma$ varies with the Fermi level $E_\mathrm{F}$, passing through $\pi$ as $E_\mathrm{F}$ traverses the energy gap that opens due to electron-hole hybridization.
We show that the evolution of $\Gamma$ is a manifestation of the pseudospin texture that encodes the momentum-dependent electron-hole coupling and is therefore a bulk signature of the nontrivial band topology.

\end{abstract}
\maketitle
Bringing the concept of topology into band theory has opened a new paradigm in classifying and distinguishing states of matter~\cite{Qi2011a}.
Unlike broken-symmetry states, which are characterized by a local quantity---the order parameter, topological phases are distinguished by an integer referred to as a topological invariant that characterizes the global properties of the band structure independently of the material details.
In two spatial dimensions, the topology of an electronic band can be characterized using an integer known as the Chern number $C$~\cite{Thouless1982, Kohmoto1985, Niu1985, Sheng2006}, which is defined for a given band as the integral of the Berry curvature over the first Brillouin zone, giving its contribution to the transverse conductivity in units of $e^{2}/h$.
($e$ is the elementary charge and $h$ is Planck's constant.)
Bulk-edge correspondence prescribes that the number of gapless edge modes equals the total Chern number of the occupied bands in the bulk.
Two-dimensional (2D) time-reversal invariant topological insulators (TIs), or quantum spin Hall insulators, characterized by their helical edge modes with up and down spins moving in opposite directions, can be thought of as a superposition of two integer quantum Hall (QH) systems with $C = \pm 1$ that are time-reversal counterparts of each other.
Experimental studies aiming to establish the hallmarks of 2D TIs in semiconductor quantum wells (QWs)~\cite{Konig2007, Konig2008, Knez2011, Suzuki2013} and various 2D materials~\cite{Pauly2015,Reis2017,Wu2018,Collins2018} have so far focused on their edge properties.
However, experimental signatures of the nontrivial band topology of the bulk that underlies the existence of the edge modes have not been detected.

Here we report the observation of the Berry phase that signifies the nontrivial band topology of a coupled electron-hole system in InAs/InGaSb QWs in the inverted regime.
The Berry phase, obtained from the phase offsets of Shubnikov-de Haas (SdH) oscillations and QH effects, varies from $0$ to $2\pi$ as the Fermi level $E_\mathrm{F}$ is tuned across the energy gap that forms through electron-hole hybridization.
We show that the Berry phase originates from the texture of the pseudospin that describes the electron-hole coupling under band inversion and the variation in the Berry phase with $E_\mathrm{F}$ is a manifestation of the underlying Berry curvature.

The topological nature of 2D TIs arises from the band inversion between the first electron (E1) and heavy-hole (HH1) subbands, which can be captured by the four-band effective Hamiltonian introduced by Bernevig, Hughes, and Zhang (BHZ) to predict the 2D TI phase in HgTe/CdTe
QWs~\cite{Bernevig2006a,Konig2008,Qi2011a}.
For the basis of $\left\vert
\mathrm{E1},\uparrow\right\rangle $, $\left\vert \mathrm{HH1},\uparrow
\right\rangle $, $\left\vert \mathrm{E1},\downarrow\right\rangle $, and
$\left\vert \mathrm{HH1},\downarrow\right\rangle $ ($\uparrow$ and $\downarrow$ represent the spin up and down states, respectively), it takes the form
\begin{align}
\label{hamiltonian}
\mathcal{H}(\mathbf{k})  &  =\left[
\begin{array}
[c]{cc}%
h(\mathbf{k}) & 0\\
0 & h^{\ast}(-\mathbf{k})
\end{array}
\right] \\
h(\mathbf{k})  &  =\epsilon(k)\mathbb{I}_{2\times2}+\mathbf{d}(\mathbf{k})\cdot\mathbf{\sigma},\nonumber
\end{align}
where $\mathbb{I}_{2\times2}$ is the $2\times2$ identity matrix and
$\epsilon(k)= -\mathcal{D}k^{2}$ with $k=|\mathbf{k}|$ and $\mathbf{k}=(k_{x},k_{y})$.
$\mathbf{d}(\mathbf{k})=(\mathcal{A}k_{x},-\mathcal{A}k_{y},\mathcal{M}-\mathcal{B}k^{2})$ is the effective field that acts on the pseudospin describing the electron-hole orbital degree of freedom, with
$\mathbf{\sigma}=(\sigma_1, \sigma_2, \sigma_3)$ the Pauli matrices.
$\mathcal{A}$, $\mathcal{B}$, $\mathcal{D}$, and $\mathcal{M}$ are material- and layer-structure-dependent parameters that determine the band structure.
The upper and lower $2\times2$ blocks of $\mathcal{H}(\mathbf{k})$ are related by time-reversal symmetry, thus yielding twofold degenerate energy bands
\begin{align}
E_{\pm}(\mathbf{k})  &  =\epsilon(k)\pm|\mathbf{d}(\mathbf{k})|\nonumber\\
&  = -\mathcal{D}k^{2}\pm\sqrt{\mathcal{A}^{2}k^{2}+(\mathcal{M}%
-\mathcal{B}k^{2})^{2}}
\label{dispersion}
\end{align}
for a system with inversion symmetry.

\begin{figure}[ptb]
\includegraphics[width = 0.48\textwidth]{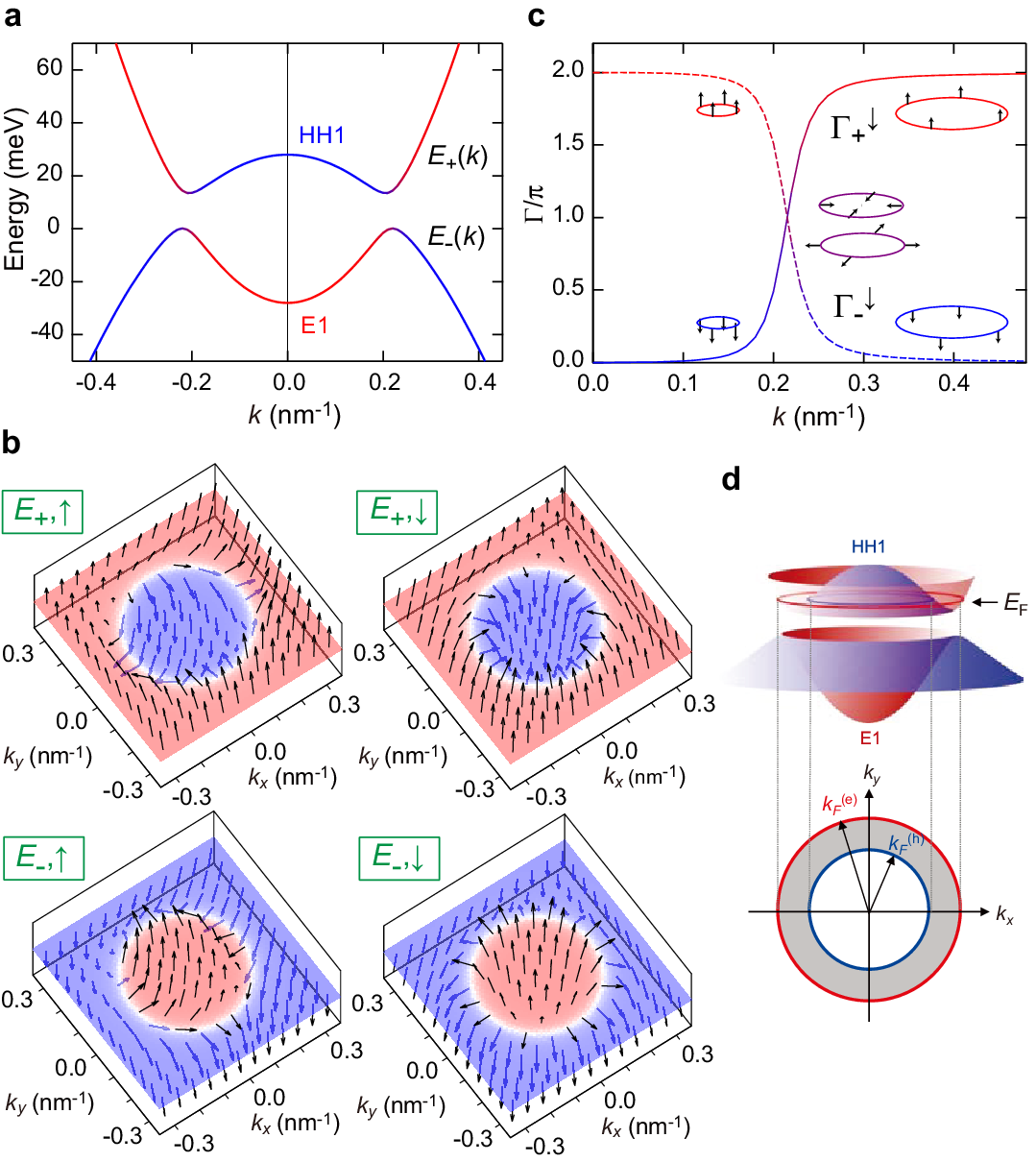}
\caption{
\textbf{Pseudospin texture and Berry phase in the BHZ model.}
\textbf{a.}
Energy dispersion with inverted band ordering calculated using equation~(\ref{dispersion}).
Parameters are $\mathcal{A} = 0.32$ eV\AA , $\mathcal{B} = -60$ eV\AA $^{2}$, $\mathcal{D} = -15$ eV\AA $^{2}$, and $\mathcal{M} = -0.028$ eV.
The red (blue) color indicates the character of the wave functions as being electron-like (hole-like).
\textbf{b}
Direction of pseudospin at each point in momentum space for the upper and lower bands with spin up and down.
The red (blue) background represents regions where the pseudospin points upwards (downwards).
\textbf{c.}
Berry's phase $\Gamma$ vs $k$ calculated using equation~(\ref{berry phase}) for the upper and lower bands with spin down $E_\pm, \downarrow$.
Insets depict the pseudospin direction on the Fermi contour and its evolution with $k$.
\textbf{d.}
Three-dimensional representation of the energy bands shown in \textbf{a} and the Fermi contours for electron-like and
hole-like quasiparticles. }
\end{figure}

Figure~1a shows an example of energy dispersion with the inverted band ordering that occurs for $\mathcal{M}\mathcal{B}>0$~\cite{Qi2011a}.
The nontrivial topology embedded in these bands can be visualized by plotting the direction of the pseudospin in momentum space~\cite{Qi2006} (Fig.~1b).
Upon moving away from $k=0$, the pseudospin gradually changes direction from down to up (or up to down), taking a vortex-like configuration at $k=\sqrt{\mathcal{M}/\mathcal{B}}$ where the out-of-plane component of the effective field $d_{3}(k)$ ($=\mathcal{M}-\mathcal{B}k^{2}$) vanishes and changes sign.
The lower bands $E_{-}(\mathbf{k})$ with up and down spins have Chern numbers $C_{\uparrow}=+1$ and $C_{\downarrow}=-1$, respectively, which underlie the quantum spin Hall effect that would occur if $E_\mathrm{F}$ is in the gap (i.e., if only these bands are filled)~\cite{Qi2011a}.

Our focus is the case where $E_\mathrm{F}$ is tuned slightly away from the gap so that quasiparticles carry current through the bulk.
Under a perpendicular field, the quasiparticles move along the Fermi contours in momentum space (Fig.~1d).
When completing a closed loop, they acquire a Berry phase $\Gamma$ proportional to the Berry curvature integrated over the area enclosed by the loop.
$\Gamma$ equals half the solid angle subtended by the pseudospin while the quasiparticles go around the loop, and is given by~\cite{Krueckl2012}
\begin{equation}
\label{berry phase}
\Gamma_{\pm}^{\uparrow}=\pi\left[  1\pm\frac{d_{3}(k)}{|\mathbf{d}(\mathbf{k})|}\right]
=\pi\left[  1\pm\frac{\mathcal{M}-\mathcal{B}k^{2}}%
{\sqrt{\mathcal{A}k^{2}+(\mathcal{M}-\mathcal{B}k^{2})^{2}}}\right],
\end{equation}
for the upper ($+$) and lower ($-$) bands with spin up.
For spin down, $\Gamma_{\pm}^{\downarrow} = \Gamma_{\mp}^{\uparrow}$.
Reflecting the momentum-space topology of the pseudospin, $\Gamma_{\pm}^{\uparrow, \downarrow}$ varies from $0$ to $2\pi$ as a function of $k$, passing through $\pi$ at $k=\sqrt{\mathcal{M}/\mathcal{B}}$ (Fig.~1c).
We emphasize that, while  $\Gamma_{\pm}^{\uparrow, \downarrow}$ can generally take non-zero values, it passes through $\pi$ only in the case of inverted bands~\cite{Krueckl2012}.
It is this phase evolution that we demonstrate in this study as a hallmark of the nontrivial band topology in the bulk.

The system we study is a semiconductor QW comprising 10-nm-thick InAs and 6-nm-thick In$_{0.25}$Ga$_{0.75}$Sb sandwiched between AlSb barriers, characterized by its ``type-II'' band alignment with the conduction band bottom of InAs located below the valence band top of In$_{0.25}$Ga$_{0.75}$Sb (Fig.~2a, upper panel)~\cite{Akiho2016,Du2017}.
For the layer thicknesses used here, the system is in the inverted regime; the lowest electron level (E1) is below the highest heavy-hole level (HH1) at $k=0$, where electrons and holes are confined separately in the InAs and In$_{0.25}$Ga$_{0.75}$Sb layers (Fig.~2a, lower panel).
The off-diagonal terms $\propto \mathcal{A}(k_{x}\pm \mathrm{i} k_{y})$ in equation (1) mix the electron and hole wave functions at finite $k$, thereby opening a hybridization gap between $E_{+}(\mathbf{k})$ and $E_{-}(\mathbf{k})$.
When $|\mathcal{A}| <\sqrt{2(\mathcal{D}-\mathcal{B})|\mathcal{M}|}$ as in the InAs/(In)GaSb system, the gap minimum occurs at a finite $k$ ($\sim \sqrt{\mathcal{M}/\mathcal{B}}$) (Fig.~1a)~\cite{Liu2008}.
This results in two concentric Fermi circles (Fig.~1d), generating electron-like and hole-like quasiparticles which coexist over a range of $E_\text{F}$ near the hybridization gap.
We tune $E_\mathrm{F}$, and hence the electron and hole densities ($n_\text{e}$ and $n_\text{h}$), by using the front-gate voltage $V_\mathrm{g}$.

\begin{figure*}[ptbh]
\includegraphics[width = 1.0\textwidth]{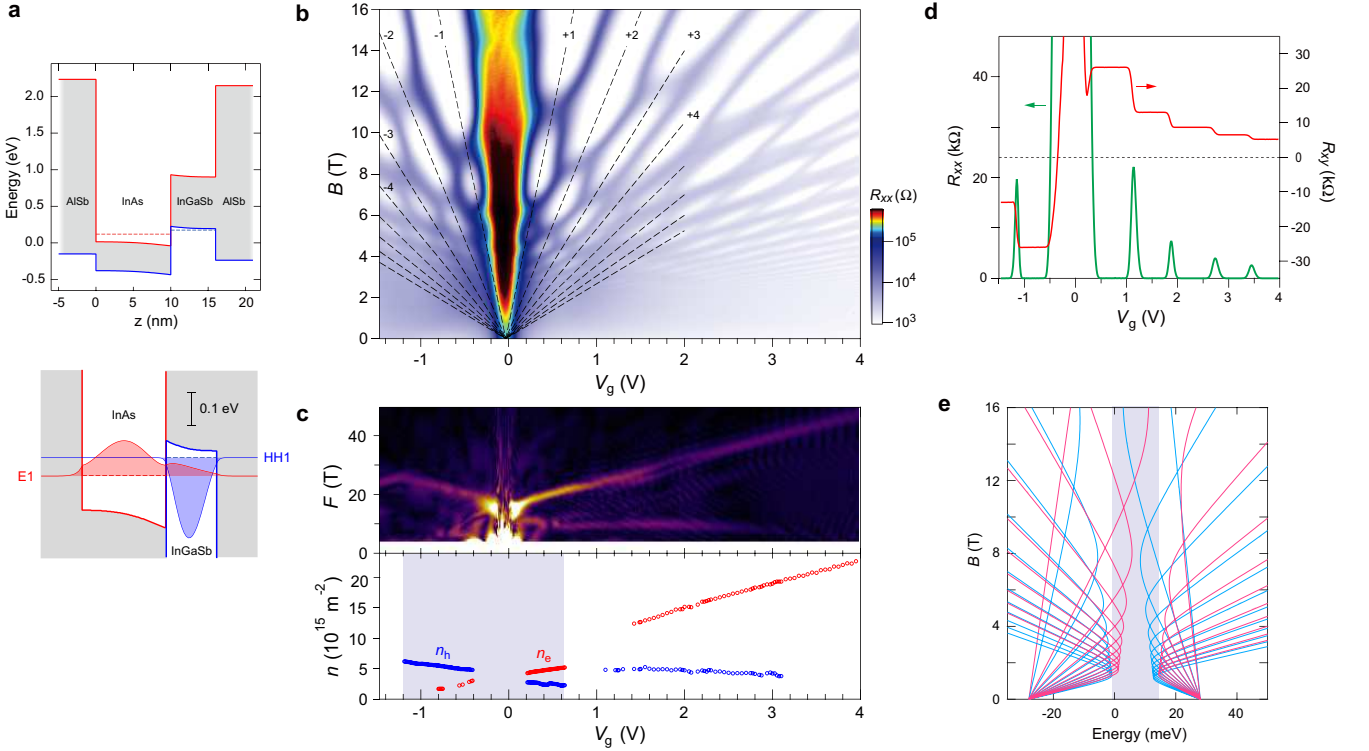}
\caption{
\textbf{Magneto-transport in the InAs/InGaSb QW.}
\textbf{a.}
(upper panel) Band-edge profile of AlSb/InAs/In$_{0.25} $Ga$_{0.75}$Sb/AlSb QW used in this study, obtained from a self-consistent $\mathbf{k}\cdot\mathbf{p}$ calculation.
(lower panel) Close-up of the band-edge profile near the Fermi level and the square of the envelope functions for the E1 and HH1 levels at $k=0$.
 \textbf{b.}
Mapping of longitudinal resistance $R_{xx}$ in the gate-voltage ($V_\mathrm{g}$)--magnetic-field ($B$) plane measured at $20$~mK.
Dashed lines represent the positions where the net filling factor $\nu_\text{net} = (n_\text{e} - n_\text{h})h/eB$ takes integer values.
\textbf{c.}
(upper panel) Fourier-transform power spectrum of the $R_{xx}$ vs $1/B$ data in \textbf{b}.
(lower panel) Density of electrons and holes obtained from the power spectrum.
\textbf{d.}
Hall resistance $R_{xy}$ and longitudinal resistance $R_{xx}$ at $B=16$~T.
\textbf{e.}
Landau level energy spectrum calculated using the BHZ effective
Hamiltonian. Landau levels belonging to the spin-up (spin-down) sector are
shown in magenta (cyan).
}
\end{figure*}

Figure 2b shows the longitudinal resistance $R_{xx}$ measured at 20~mK as a function of $V_\mathrm{g}$ and magnetic field $B$ applied perpendicular to the sample.
The large resistance peak at $V_\mathrm{g}\sim0$~V corresponds to the charge neutrality point (CNP), where $E_\text{F}$ traverses the gap and $n_{\text{e}}$ and $n_{\text{h}}$ become equal.
$R_{xx}$ oscillates with $B$ and $V_\mathrm{g}$ in a complex manner, reflecting the coexistence of electrons and holes.
With increasing $B$, the $R_{xx}$ minima deepen and develop into wide QH regions with vanishing $R_{xx}$ and quantized Hall resistance $R_{xy}$ (Fig.~2d).
In previous studies on electron-hole systems in InAs/GaSb QWs, QH effects were observed when the filling factors of the electron and hole Landau levels (LLs), $\nu_{j}=n_{j}h/eB$ ($j=$ e, h), were both integers~\cite{Mendez1985}. 
The Hall conductance $\sigma_{xy}$ was thus quantized to $\nu_{\mathrm{net}}e^{2}/h$, with $\nu_{\mathrm{net}}=n_{\mathrm{net}}h/eB=(n_{\mathrm{e}}-n_{\mathrm{h}})h/eB$ the net filling factor.
Here, we obtain $n_{\mathrm{e}}$ and $n_{\mathrm{h}}$ (and hence $n_\text{net}$) as a function of $V_\mathrm{g}$ from the Fourier power spectra of $R_{xx}$ vs $1/B$ (Fig.~2c)~\cite{Akiho2016} (see Supplementary Information for details).
The data in Fig.~2b confirm that the QH regions at high fields as well as the $R_{xx}$ minima at low fields appear when $\nu_{\mathrm{net}}$ becomes an integer (shown by dashed lines).

Examining the filling factors of the individual carriers, however, reveals nontrivial behavior.
Figure~3 shows the  $(V_\mathrm{g}, B)$ map of the longitudinal conductivity $\sigma_{xx}$ near the CNP, calculated from  the measured $R_{xx}$ and $R_{xy}$.
The red (blue) lines represent the field at which $\nu_{\mathrm{e}}$ ($\nu_{\mathrm{h}}$), as calculated from $n_\mathrm{e}$ ($n_\mathrm{h}$), becomes an integer at each $V_\mathrm{g}$ (see Supplementary Information for details).
The data reveal that the positions of the $\sigma_{xx}$ minima and QH regions (indicated by black dots) significantly deviate from the fields at which $\nu_\mathrm{e}$ or $\nu_\mathrm{h}$ becomes an integer.
This clearly shows that near the CNP each carrier acquires a large nonzero Berry phase.
It is important to note that, for a given $E_\text{F}$ near the CNP ($-1.2 \leq V_\mathrm{g} \lesssim 0.63$~V, shaded area in Fig.~2c), only LLs from one spin sector are involved in transport~\cite{Akiho2016}.
This happens because electron and hole LLs hybridize in different manners for the up and down spins, leaving only spin-down (spin-up) LLs right above (below) the hybridization gap (Fig.~2e)~\cite{Zhang2014}.
This allows us to ignore the spin degree of freedom in our discussion.

\begin{figure}[t]
\includegraphics[width = 0.45\textwidth]{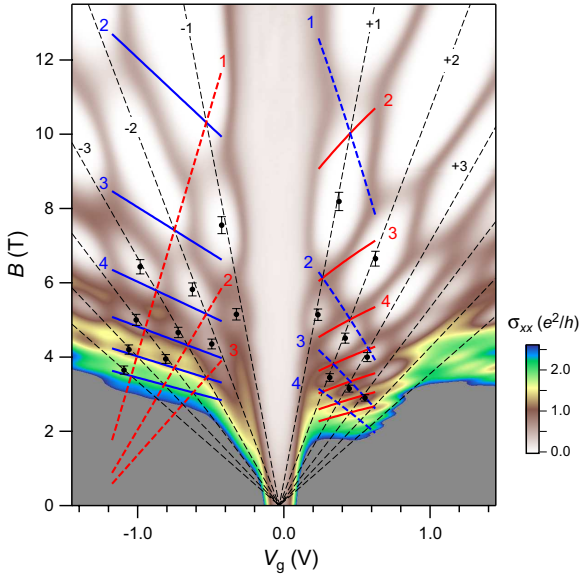}
\caption{
\textbf{Comparison of $\sigma_{xx}$ minima positions and expected filling factors of electron and hole Landau levels.}
Color plot of $\sigma_{xx}$ in the $V_\mathrm{g}$--$B$ plane.
Regions with $\sigma_{xx} > 2.6 e^2/h$ are shown in gray.
Black dots indicate the positions of $\sigma_{xx}$ minima near the CNP, where the error bars represent $\pm 3$\% uncertainty in $B$.
The red (blue) lines represent the positions where the filling factor of electron (hole) LLs calculated from the SdH frequency at each $V_\mathrm{g}$ becomes an integer for the case of $\Gamma = 0$.
Filling factors of the majority (minority) carrier are shown as solid (dashed) lines.
}
\end{figure}

As the $R_{xx}$ oscillations in Fig.~2b contain two frequencies associated with the electron-like and hole-like Fermi contours whose areas do not differ significantly, the standard method to determine Berry's phase from the field positions of oscillation minima at a given $V_\mathrm{g}$ cannot be used.
We therefore focus on each point of $\sigma_{xx}$ minima in the $(V_{\mathrm{g}},B)$ map, which corresponds to oscillation minima for both carriers.
Using the field position $B_{\mathrm{min}}$ of the $\sigma_{xx}$ minimum and the SdH oscillation frequency $F^{(j)}$ for each carrier ($j=$ e, h) at the relevant $V_\mathrm{g}$, Berry's phase can be deduced as $\Gamma^{(j)}=2\pi\left[  N^{(j)}-\frac{F^{(j)}}{B_{\mathrm{min}}}\right]  $, where $N^{(j)}$ is an integer which we chose so that $0\leq\Gamma^{(j)} < 2\pi$ (see Supplementary Information for details).
The $\Gamma^{(j)}$ values extracted from several $\sigma_{xx}$ minima at $-1.2 \leq V_\mathrm{g} \leq 0.63$~V are plotted as a
function of $V_\mathrm{g}$ (Fig.~4a) and Fermi wave number $k_{F}^{(j)}=\sqrt{S^{(j)}/\pi}$ (Fig.~4b), where $S^{(j)}=\frac{e}{h}F^{(j)}$ is the Fermi contour area.
Note that in a magnetic field quasiparticles belonging to hole-like orbits enclosing states with higher energies move in the opposite direction in momentum space~\cite{Ashcroft1976} (Fig.~4a, inset).
We therefore took $\Gamma^{(\mathrm{h})}= -\Gamma^{(\mathrm{e})}$  ($\bmod \, 2\pi$) for the theoretical curve in Fig.~4b.
Our analysis reveals that the measured $\Gamma^{(j)}$ varies between $0$ and $2\pi$, passing through $\pi$, as predicted by the BHZ model.
We note that the $\pi$ Berry phase observed near the CNP is consistent with the recent theory of de Haas-van Alphen effect in narrow-gap topological insulators~\cite{Zhang2016}, which, using a similar Hamiltonian, predicts a $\pi$ Berry phase for in-gap oscillations.

\begin{figure}[ptb]
\includegraphics[width = 0.40\textwidth]{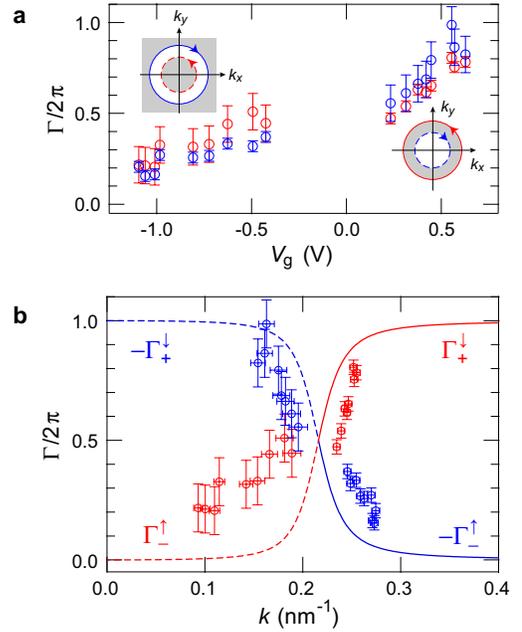}
\caption{
\textbf{Evolution of the Berry phase across the hybridization gap.}
\textbf{a.}
Berry phase $\Gamma$ deduced from the field position of $\sigma_{xx}$ minima and the SdH frequency at each $V_\mathrm{g}$.
The data shown in red (blue) were obtained from the SdH frequency corresponding to the Fermi circle area of electrons (holes) at each $V_\mathrm{g}$.
Inset: Fermi circles of electrons (red) and holes (blue) at $V_\mathrm{g} < 0$ (left) and $V_\mathrm{g} > 0$ (right).
Arrows indicate the direction of motion in a perpendicular magnetic field.
\textbf{b.}
$\Gamma$ values in \textbf{a} replotted as a function of Fermi wave number $k$.
The red and blue lines show the Berry phase $\Gamma^\text{(e,h)}$ ($\bmod \, 2\pi$) for electrons and holes, respectively, calculated from the BHZ model [equation~(\ref{berry phase})].
The suffixes of $\Gamma_{\pm}^{\uparrow, \downarrow}$ indicate the band index ($\pm$) and spin ($\uparrow, \downarrow$) of the states relevant to the magnetotransport at each $k$.
See Supplementary Information for the estimation of the error.
}
\end{figure}

A caveat should be mentioned here.
Our analysis is based on a Hamiltonian in which the spin-orbit coupling between the upper and lower $2\times2$ blocks is neglected.
In reality, spin-orbit coupling exists, which at $B=0$ lifts the twofold degeneracy away from $k=0$ due to the structural inversion asymmetry~\cite{Liu2008}.
However, as shown in Fig.~2e, due to the spin-dependent electron-hole coupling, LLs from different spin sectors are energetically well separated around the CNP, which makes our analysis a reasonably good approximation.

It is essential that the Berry phase observed here is associated with pseudospin and not real spin.
Since real spins are tilted toward the magnetic field by the Zeeman coupling, the associated Berry phase is necessarily reduced below $\pi$.
Recently, a nonzero Berry phase of $\Gamma=0.3\pi$ was reported for a similar electron-hole system in InAs/GaSb QWs~\cite{Nichele2017}, which was discussed in terms of spin-momentum locking due to the Rashba-type spin-orbit coupling.
Our data revealing the variation in the Berry phase as a function of gate voltage around the CNP (Fig.~4) demonstrate that it originates from the winding of a pseudospin and not a real spin.
We add that field-induced interlayer charge transfer, which may cause a phase slip in the SdH oscillations away from the CNP~\cite{Karalic2019}, is not relevant near the CNP.

In a previous study, a $\pi$ Berry phase has been reported for a HgTe/CdTe QW at the critical thickness designed to mimic massless Dirac fermions in graphene~\cite{Buttner2011}.
However, the continuous evolution of $\Gamma$ as a function of $E_\text{F}$, expected for HgTe/CdTe QWs in the inverted regime~\cite{Krueckl2012}, has not been observed.
We note that, in HgTe/CdTe QWs, both spin-up and spin-down Fermi surfaces with nearly the same area are always involved in transport.
Since the spin-up and spin-down quasiparticles acquire opposite Berry phases, the resultant phase shifts only appear as a change in the relative strength of even and odd integer fillings~\cite{Buttner2011}; this makes it unfeasible to extract $\Gamma$ for each spin component from SdH oscillations.
We can thus understand that the detection of the non-trivial Berry phase in InAs/(In)GaSb QWs is made possible by their special Landau-level structure in which only a single spin species resides at the Fermi level near the CNP.
Finally, we should point out that the Berry phase is proportional to the Berry curvature integrated over the Fermi circle.
The variation in $\Gamma$ as a function of $E_\mathrm{F}$ observed around the CNP is thus a direct manifestation of a large Berry curvature that exists near the CNP.
Our results, showing a way to probe and control the Berry curvature in situ, would be useful in engineering various topological systems.

\subsection{Methods}

The heterostructure was grown by molecular-beam epitaxy on a Si-doped GaAs (001) substrate.
The QW consisted of a $10$-nm InAs top layer and a $6$-nm In$_{0.25}$Ga$_{0.75}$Sb bottom layer, grown pseudomorphically on a fully relaxed 800-nm-thick AlSb buffer layer, and capped with 50-nm AlSb and 5-nm GaSb.
The sample was processed into a $50$-$\mu$m-wide Hall bar with voltage-probe distance of $180$~$\mu$m.
Ohmic contacts were made after etching down to the InAs layer and depositing Ti/Au and lift off.
The Hall bar was fitted with a Ti/Au front gate fabricated on an atomic-layer-deposited 40-nm-thick Al$_{2}$O$_{3}$ insulator.
In$_{0.25}$Ga$_{0.75}$Sb has a lattice constant 0.82\% larger than that of AlSb, which induces in-plane compressive strain in the InGaSb
layer.
As shown Ref.~\cite{Akiho2016}, a larger hybridization gap can be obtained by virtue of the strain-induced band engineering.
Transport measurements were performed using a lock-in technique with an excitation current of 1-20~nA.

\paragraph{\textbf{{Acknowledgements}}}

The authors thank H. Murofushi for help during the sample processing.
K.M. thanks Kentaro Nomura and Makoto Kohda and F.C. thanks Jean-No\"{e}l Fuchs for useful discussions.
This work was supported by JSPS KAKENHI Grant Number JP15H05854.

\paragraph{\textbf{{Author contributions}}}

F.C., H.I., and T.A. performed the transport measurements.
F.C., H.I., and K.M. analyzed the data and wrote the manuscript.
T.A. and K.O. grew the heterostructure, H.I. processed the samples, and F.C., H.I., T.A., and K.S. characterized the samples.
All authors discussed the results and commented on the manuscript.

\paragraph{\textbf{{Additional information}}}

Correspondence and requests for materials should be addressed to K.M.

\paragraph{\textbf{{Competing financial interests}}}

The authors declare no competing financial interests.


\end{document}